\documentclass[notitlepage,aps,prd,a4paper,11pt,amsfonts,amssymb,amsmath]{revtex4-1}
\usepackage{enumerate}
\usepackage[all]{xy}

\newcommand{\mat}[1]{\underline{\underline{#1}}}
\newcommand{\vek}[1]{\underline{#1}}

\begin{document}
\title{Propagation of gravitational waves in multimetric gravity}

\author{Manuel Hohmann}
\email{manuel.hohmann@desy.de}
\affiliation{Zentrum f\"ur Mathematische Physik und II. Institut f\"ur Theoretische Physik, Universit\"at Hamburg, Luruper Chaussee 149, 22761 Hamburg, Germany}

\begin{abstract}
We discuss the propagation of gravitational waves in a recently discussed class of theories containing \(N \geq 2\) metric tensors and a corresponding number of standard model copies. Using the formalism of gauge-invariant linear perturbation theory we show that all gravitational waves propagate at the speed of light. We then employ the Newman-Penrose formalism to show that two to six polarizations of gravitational waves may exist, depending on the parameters entering the field equations. This corresponds to E(2) representations \(\mathrm{N}_2\), \(\mathrm{N}_3\), \(\mathrm{III}_5\) and \(\mathrm{II}_6\). We finally apply our general discussion to a recently presented concrete multimetric gravity model and show that it is of class \(\mathrm{N}_2\), i.e., it allows only two tensor polarizations, as it is the case for general relativity. Our results provide the theoretical background for tests of multimetric gravity theories using the upcoming gravitational wave experiments.
\end{abstract}
\maketitle

\section{Motivation}\label{sec:motivation}
This article continues a series of articles~\cite{Hohmann:2009bi,Hohmann:2010vt,Hohmann:2010ni} discussing gravity theories with \(N \geq 2\) metric tensors \(g^I\) and a corresponding number of standard model copies \(\Psi^I\), where each copy of the standard model couples only to its own metric tensor and the interaction between the different standard model copies is mediated solely by an interaction between the different metrics. We are particularly interested in theories that exhibit repulsive gravitational forces between different standard model copies in the Newtonian limit which are of equal strength compared to the attractive gravitational force within each standard model copy. While this is not possible~\cite{Hohmann:2009bi} for \(N = 2\), it may serve as a potential explanation for the observed late-time acceleration of the universe~\cite{Hohmann:2010vt} for \(N \geq 3\) and is consistent with high-precision solar system experiments at the post-Newtonian level~\cite{Hohmann:2010ni}. The theories we consider satisfy the following assumptions:

\begin{enumerate}[\it (i)]
\item\label{ass:action}
\textit{The action is of the form}
\begin{equation}\label{eqn:actionsplit}
S = S_G[g^1, \ldots, g^N] + \sum_{I = 1}^{N}S_M[g^I, \Psi^I]\,,
\end{equation}
\textit{where \(S_G\) is the gravitational part of the action and \(S_M\) is the standard model action.}\\
This assumption guarantees that each standard model copy \(\Psi^I\) couples only to its own metric tensor \(g^I\), so that the standard model fields satisfy the same field equations as they would in a single-metric theory. It further excludes any non-gravitational interaction between the different standard model copies, which implies that the different standard model copies appear mutually dark.

\item\label{ass:tensor}
\textit{The gravitational field equations are obtained by variation with respect to the metrics~\({g^1_{ab} \ldots g^N_{ab}}\), and so are a set of symmetric two-tensor equations of the form \(\vek{K}_{ab} = 8\pi G_N\vek{T}_{ab}\).}\\
This employs the well-known principle of stationary action. It follows that the number of field equations equals the number of field components.

\item\label{ass:derivatives}
\textit{The geometry tensor \(\vek{K}_{ab}\) contains at most second derivatives of the metrics, which can be achieved by a suitable choice of the gravitational action.}\\
This assumption is one of mathematical simplicity and guarantees a reasonable amount of technical control over the partial differential field equations. It will be used to restrict the possible terms in the linearized field equations of our theory.

\item\label{ass:symmetry}
\textit{The field equations are symmetric with respect to arbitrary permutations of the sectors~\((g^I, \Psi^I)\).}\\
This is another assumption made for simplicity; it employs the Copernican principle in the sense that the same laws of nature should hold within each sector. It also follows that the interaction between the different sectors will satisfy Newton's principle that action equals reaction for the gravitational forces in the Newtonian limit.

\item\label{ass:flat}
\textit{The vacuum solution is given by a set of flat metrics \(g^I_{ab} = \eta_{ab}\).}\\
Cosmological constants are excluded because we are interested in multimetric gravity theories in which the accelerating universe is modelled by a repulsive interaction between different standard model copies, as we have shown in~\cite{Hohmann:2010vt}. Further assuming a simultaneous maximal set of Killing symmetries for all metrics yields the stated vacuum solution.
\end{enumerate}

The aim of this article is to examine the propagation of gravitational waves in the weak field limit of multimetric gravity theories satisfying the aforementioned assumptions. Note that in addition to the recently discussed theories with \(N \geq 3\) this also includes bimetric gravity theories such as~\cite{arXiv:0912.0790,arXiv:1001.4444,arXiv:1006.3809}; see \cite{arXiv:1006.5619} for an overview of bimetric theories and a discussion of their weak field limits. In particular, we aim to calculate two properties of gravitational waves which are expected to be accessible by the upcoming detector experiments. One of these properties is the polarization of gravitational waves. Metric gravity theories can be classified by the presence of up to six polarizations in terms of representations of the little group E(2), which may be distinguished by measuring the electric components \(R_{0\alpha 0\beta}\) of the Riemann tensor~\cite{Eardley:1973br,Eardley:1974nw}. Although gravitational waves have not been observed yet, the sensitivity of present and future experiments is continuously being improved, and it is expected that a sensitivity sufficient for the detection of gravitational waves will be reached within the next years, hence providing a valuable instrument for testing gravity theories~\cite{Thorne:1987af,Sathyaprakash:2009xs}.

The other property is the propagation velocity \(v_g\), which equals the speed of light in general relativity, but may differ significantly in multimetric gravity theories containing massive gravitons~\cite{Berezhiani:2007zf,Berezhiani:2009kv}. Theories of this type may be tested by comparing the arrival times of gravitational radiation and light from distant supernovae~\cite{Will:1993ns}, taking into account the possibility of a different Shapiro delay for both types of radiation~\cite{Desai:2008vj,Kahya:2010dk}. Apart from the direct observation of gravitational waves, bounds on \(v_g\) can also be obtained through indirect observations. An upper bound on \(c - v_g\) is placed by the observation of high-energy cosmic rays: massive particles whose velocity exceeds \(v_g\) should be decelerated due to the emission of gravitational bremsstrahlung~\cite{Caves:1980jn}. Another bound on \(c - v_g\) can be obtained from pulsar timing: if \(v_g < c\), the interaction between electromagnetic and gravitational radiation should influence the arrival times of radio signals in a gravitational wave background~\cite{Polnarev:2008tz,Baskaran:2008za}.

A complete calculation of the aforementioned effects such as the Shapiro delay of both light and gravitational waves in cosmic gravitational fields or the mutual interaction between both types of radiation requires a treatment based on the full non-linear field equations of a concrete multimetric gravity theory. Since it is the aim of this article to derive the general properties of gravitational waves for a large class of multimetric gravity theories, we will not perform this calculation here. Instead, we will make use of assumption~\textit{(\ref{ass:flat})} and consider the propagation of gravitational waves in a flat background metric. We then compare their propagation velocity \(v_g\) to the fundamental velocity \(c\), which equals the physical speed of light in this background according to assumption~\textit{(\ref{ass:action})}.

The outline of this article is as follows. In section~\ref{sec:velocity} we will determine the propagation velocity of gravitational waves in a flat Minkowski background. For this purpose we will apply the gauge invariant linear perturbation formalism known from cosmological perturbation analysis~\cite{Bardeen:1980kt,Malik:2008im,Stewart:1990fm} to the most general linearized vacuum field equations of multimetric gravity in subsection~\ref{subsec:gaugeinv}. This formalism allows us to separate the physical degrees of freedom from pure gauge quantities. In subsection~\ref{subsec:bianchi} we will discuss the role of the Bianchi identities in the class of theories we consider. We will then calculate the wave-like solutions of the linearized vacuum field equations in subsection~\ref{subsec:wave} and show that all wave-like solutions are null waves. In section~\ref{sec:newpen} we will use the Newman-Penrose formalism~\cite{Newman:1961qr} to determine the allowed polarizations. The findings of sections~\ref{sec:velocity} and~\ref{sec:newpen} will then be applied to two concrete example theories in section~\ref{sec:examples}. We will conclude with a discussion in section~\ref{sec:conclusion}.

\section{Propagation velocity}\label{sec:velocity}
In this section we will calculate the propagation velocity \(v_g\) of gravitational waves within the class of multimetric theories satisfying assumptions~\textit{(\ref{ass:action})} to~\textit{(\ref{ass:flat})} stated in the introduction. The starting point of our calculation will be a perturbation ansatz around the vacuum solution, which is a set of flat Minkowski metrics according to assumption~\textit{(\ref{ass:flat})}. This ansatz leads us to the most general linearized multimetric vacuum field equations satisfying our assumptions. We will employ the gauge-invariant formalism detailed in~\cite{Hohmann:2009bi} in order to determine the physical degrees of freedom. It will turn out that the only wave-like solutions of the gauge-invariant field equations are null waves.

From assumption~\textit{(\ref{ass:action})} it follows further that light rays constituted by the electromagnetic field of one standard model copy \(\Psi^I\) follow the lightlike geodesics of the corresponding metric \(g^I_{ab}\). The fact that we use the aforementioned perturbation ansatz allows us to conclude that these geodesics are the lightlike directions of the Minkowski background, up to higher order perturbations which we neglect. Hence, we will conclude that all gravitational waves propagate at the speed of light.

\subsection{Gauge-invariant formalism}\label{subsec:gaugeinv}
For the derivation presented in this section it is sufficient to treat gravitational waves as a small perturbation of the metrics around a vacuum solution of the field equations. Making use of assumption~\textit{(\ref{ass:flat})}, we thus use the perturbation ansatz
\begin{equation}\label{eqn:perturb}
g^I_{ab} = \eta_{ab} + h^I_{ab}\,,
\end{equation}
where the components \(h^I_{ab}\) are small, \(|h^I_{ab}| \ll 1\). Under this condition the most general linearized vacuum field equations compatible with our assumptions \textit{(\ref{ass:action})}-\textit{(\ref{ass:flat})} stated in the introduction take the form~\cite{Hohmann:2009bi}
\begin{equation}\label{eqn:linvaceom}
0 = \vek{K}_{ab} = \mat{P} \cdot \partial^p\partial_{(a}\vek{h}_{b)p} + \mat{Q} \cdot \square\vek{h}_{ab} + \mat{R} \cdot \partial_a\partial_b\vek{h} + \mat{M} \cdot \partial^p\partial^q\vek{h}_{pq}\eta_{ab} + \mat{N} \cdot \square\vek{h}\eta_{ab}\,,
\end{equation}
where indices are raised with the flat metric \(\eta\) and \(\square = \eta^{ab}\partial_a\partial_b\). The matrices \(\mat{P}, \mat{Q}, \mat{R}, \mat{M}, \mat{N}\) are constant parameters. Note that these parameter matrices are not completely arbitrary within the class of theories we consider, but are further restricted by our assumptions, as we will show in the remainder of this section. It will turn out that our assumptions do not uniquely fix the parameter matrices. For any concrete multimetric gravity theory, their values depend on the choice of the gravitational action \(S_G\) introduced in equation~\eqref{eqn:actionsplit}, and can be calculated by an explicit linearization of the full nonlinear field equations. We will give the values of the parameter matrices for two example theories in section~\ref{sec:examples}.

We now apply the gauge invariant linear perturbation formalism known from cosmological perturbation analysis~\cite{Bardeen:1980kt,Malik:2008im,Stewart:1990fm} to the linearized field equations~\eqref{eqn:linvaceom} in order to determine the physical degrees of freedom. We only sketch the procedure here; see~\cite{Hohmann:2009bi} for full detail. First, we perform a purely algebraic $(1 + 3)$ split of the spacetime coordinates \(x^a = (x^0, x^{\alpha})\) into time and space, and correspondingly decompose the metric perturbations \(\vek{h}_{ab}\) and the geometry tensors \(\vek{K}_{ab}\). Second, we perform a differential decomposition of the metric perturbations,
\begin{equation}\label{eqn:diffdecom}
\vek{h}_{00} = -2\vek{\phi}, \qquad
\vek{h}_{0\alpha} = \partial_{\alpha}\vek{\tilde{B}} + \vek{\tilde{B}}_{\alpha}, \qquad
\vek{h}_{\alpha\beta} = -2\vek{\psi}\delta_{\alpha\beta} + 2\triangle_{\alpha\beta}\vek{\tilde{E}} + 4\partial_{(\alpha}\vek{\tilde{E}}_{\beta)} + 2\vek{\tilde{E}}_{\alpha\beta}\,,
\end{equation}
into four scalars \(\vek{\phi}, \vek{\psi}, \vek{\tilde{B}}, \vek{\tilde{E}}\), two divergence-free (or transverse) vectors \(\vek{\tilde{B}}_{\alpha}, \vek{\tilde{E}}_{\alpha}\) and one divergence-free, trace-free tensor \(\vek{\tilde{E}}_{\alpha\beta}\). Here \(\triangle_{\alpha\beta} = \partial_{\alpha}\partial_{\beta} - \frac{1}{3}\delta_{\alpha\beta}\triangle\) denotes the trace-free second derivative. A similar decomposition of the geometry tensor,
\begin{equation}
\vek{K}_{00}, \qquad
\vek{K}_{0\alpha} = \partial_{\alpha}\vek{\tilde{W}} + \vek{\tilde{W}}_{\alpha}, \qquad
\vek{K}_{\alpha\beta} = \frac{1}{3}\vek{Z}\delta_{\alpha\beta} + \triangle_{\alpha\beta}\vek{\tilde{Z}} + 2\partial_{(\alpha}\vek{\tilde{Z}}_{\beta)} + \vek{\tilde{Z}}_{\alpha\beta}\,,
\end{equation}
shows that its scalar components \(\vek{K}_{00}, \vek{\tilde{W}}, \vek{Z}, \vek{\tilde{Z}}\) depend only on scalar components of the metrics, its transverse vector components \(\vek{\tilde{W}}_{\alpha}, \vek{\tilde{Z}}_{\alpha}\) depend only on vector components, and its transverse trace-free tensor components \(\vek{\tilde{Z}}_{\alpha\beta}\) depend only on tensor components. In other words, the scalar, vector and tensor components of the field equations decouple. In the next step we replace the components of the metric perturbations by the potentials
\begin{gather}
\vek{I_1} = \vek{\phi} + \partial_0\vek{\tilde{B}} - \partial_0^2\vek{\tilde{E}}\,, \qquad
\vek{I_2} = \vek{\psi} + \frac{1}{3}\triangle\vek{\tilde{E}}\,, \qquad
\vek{I_3} = \vek{\tilde{B}}\,, \qquad
\vek{I_4} = \vek{\tilde{E}}\,,\nonumber\\
\vek{I}_\alpha = \vek{\tilde{B}}_{\alpha} - 2\partial_0\vek{\tilde{E}}_{\alpha}\,, \qquad
\vek{I'}_{\alpha} = \vek{\tilde{E}}_{\alpha}\,, \qquad
\vek{I}_{\alpha\beta} = \vek{\tilde{E}}_{\alpha\beta}\,.\label{eqn:potentials}
\end{gather}
Using these quantities, we finally obtain the scalar equations
\begin{subequations}\label{eqn:scalar}
\begin{align}
\vek{K}_{00} &= 2(\mat{P} + \mat{Q} + \mat{R} + \mat{M} + \mat{N}) \cdot \partial_0^2\vek{I_1} - 2(\mat{Q} + \mat{N}) \cdot \triangle\vek{I_1} - 6(\mat{R} + \mat{N}) \cdot \partial_0^2\vek{I_2}\nonumber\\
&\phantom{=} + 2(\mat{M} + 3\mat{N}) \cdot \triangle\vek{I_2} + 2(\mat{P} + \mat{Q} + \mat{R} + \mat{M} + \mat{N}) \cdot (-\partial_0^3\vek{I_3} + \partial_0^4\vek{I_4})\\
&\phantom{=} + (\mat{P} + 2\mat{Q} + 2\mat{M} + 2\mat{N}) \cdot \partial_0\triangle\vek{I_3} + 2(\mat{R} - \mat{Q}) \cdot \partial_0^2\triangle\vek{I_4} - 2(\mat{M} + \mat{N}) \cdot \triangle\triangle\vek{I_4}\,,\nonumber\\
\vek{\tilde{W}} &= (\mat{P} + 2\mat{R}) \cdot \partial_0\vek{I_1} - (\mat{P} + 6\mat{R}) \cdot \partial_0\vek{I_2} - \frac{1}{2}(3\mat{P} + 2\mat{Q} + 4\mat{R}) \cdot \partial_0^2\vek{I_3}\nonumber\\
&\phantom{=} + \frac{1}{2}(\mat{P} + 2\mat{Q}) \cdot \triangle\vek{I_3} + (\mat{P} + 2\mat{R}) \cdot \partial_0(\partial_0^2 + \triangle)\vek{I_4}\,,\\
\vek{Z} &= -6(\mat{M} + \mat{N}) \cdot \partial_0^2\vek{I_1} + 2(\mat{R} + 3\mat{N}) \cdot \triangle\vek{I_1} + 6(\mat{Q} + 3\mat{N}) \cdot \partial_0^2\vek{I_2} + 2(\mat{R} - \mat{Q}) \cdot \partial_0^2\triangle\vek{I_4}\nonumber\\
&\phantom{=} - 2(\mat{P} + 3\mat{Q} + 3\mat{R} + 3\mat{M} + 9\mat{N}) \cdot \triangle\vek{I_2} + 6(\mat{M} + \mat{N}) \cdot (\partial_0^3\vek{I_3} - \partial_0^4\vek{I_4})\\
&\phantom{=} - (\mat{P} + 2\mat{R} + 6\mat{M} + 6\mat{N}) \cdot \partial_0\triangle\vek{I_3} + 2(\mat{P} + \mat{Q} + \mat{R} + 3\mat{M} + 3\mat{N}) \cdot \triangle\triangle\vek{I_4}\,,\nonumber\\
\vek{\tilde{Z}} &= 2\mat{R} \cdot \vek{I_1} - 2(\mat{P} + 3\mat{R}) \cdot \vek{I_2} - (\mat{P} + 2\mat{R}) \cdot \partial_0\vek{I_3} + 2(\mat{R} - \mat{Q}) \cdot \partial_0^2\vek{I_4} + 2(\mat{P} + \mat{Q} + \mat{R}) \cdot \triangle\vek{I_4}\,,
\end{align}
\end{subequations}
the vector equations
\begin{subequations}\label{eqn:vector}
\begin{align}
\vek{\tilde W}_{\alpha} &= -\frac{1}{2}(\mat{P} + 2\mat{Q}) \cdot (\partial_0^2\vek{I}_{\alpha} - 2\partial_0\square\vek{I'}_{\alpha}) + \mat{Q} \cdot \triangle\vek{I}_{\alpha}\,,\\
\vek{\tilde Z}_{\alpha} &= -\frac{1}{2}\mat{P} \cdot \partial_0\vek{I}_{\alpha} + (\mat{P} + 2\mat{Q}) \cdot \square\vek{I'}_{\alpha}\,,
\end{align}
\end{subequations}
and the tensor equations
\begin{equation}\label{eqn:tensor}
\vek{\tilde{Z}}_{\alpha\beta} = 2\mat{Q} \cdot \square\vek{I}_{\alpha\beta}\,.
\end{equation}
The reason for this rewriting becomes apparent when we determine the physical degrees of freedom. These are linear combinations of the components of the metrics which are invariant under gauge transformations. Since the gravitational fields of our theory are a set of metric tensors, the only gauge transformations we allow are diffeomorphisms of the underlying manifold, as it is also the case in general relativity. Any such gauge transformation is generated by a vector field \(\xi\) and simultaneously changes all tensor fields \(F\) by their Lie derivatives, \(\delta_{\xi}F = \mathcal{L}_{\xi}F\). Considering the special case \(F = g^I\), we find that the components of the metric perturbations transform according to
\begin{equation}\label{eqn:gatra}
\delta_{\xi}h^I_{ab} = \partial_a\xi_b + \partial_b\xi_a\,,
\end{equation}
where \(\xi_a = \eta_{ab}\xi^b\). Note that this gauge freedom is more restrictive than it would be for a set of \(N\) independent spin-2 fields, where each field has its own set of gauge transformations~\cite{Hohmann:2010ni,vanderBij:1981ym}. Since we further wish to keep the formal structure of the perturbation ansatz~\eqref{eqn:perturb}, we only consider gauge transformations in which the components \(\xi^a\) are of order \(\mathcal{O}(h)\). Applying the decomposition from equation~\eqref{eqn:diffdecom} to the vector field \(\xi^a\),
\begin{equation}
\xi_0 = \xi, \quad \xi_{\alpha} = \partial_{\alpha}\tilde{\xi} + \tilde{\xi}_{\alpha}\,,
\end{equation}
we obtain two scalars \(\xi\) and \(\tilde{\xi}\) and one divergence-free vector \(\tilde{\xi}_{\alpha}\). These can be used to write the change of the potentials~\eqref{eqn:potentials} under gauge transformations in the form
\begin{equation}
\delta_{\xi}I_1^I = \delta_{\xi}I_2^I = 0\,, \quad \delta_{\xi}I_3^I = \partial_0\tilde{\xi} + \xi\,, \quad \delta_{\xi}I_4^I = \tilde{\xi}\,, \quad \delta_{\xi}I_{\alpha}^I = 0\,, \quad \delta_{\xi}I_{\alpha}'^I = \frac{1}{2}\tilde{\xi}_{\alpha}\,, \quad \delta_{\xi}I_{\alpha\beta}^I = 0\,.
\end{equation}
One can now immediately read off the gauge-invariant quantities \(\vek{I_1}\), \(\vek{I_2}\), \(\vek{I}_{\alpha}\) and \(\vek{I}_{\alpha\beta}\). Further, linear combinations of the form \(c_II_3^I\), \(c_II_4^I\) and \(c_II_{\alpha}'^I\) (where summation over \(I\) is implied) are gauge-invariant if and only if the sum of the coefficients \(c_I\) vanishes. It thus makes sense to consider the linearly related quantities \(\vek{\mathfrak{I}} = \mat{U} \cdot \vek{I}\), where the matrix \(\mat{U}\) is given by
\begin{equation}\label{eqn:baschange}
U^{IJ} = \begin{cases}
\frac{1}{\sqrt{N}} & \text{if } I = 1 \text{ or } J = 1\,,\\
1 + \frac{1}{\sqrt{N}} + \frac{1}{1 - \sqrt{N}} & \text{if } I = J > 1\,,\\
\frac{1}{\sqrt{N} - N} & \text{otherwise.}
\end{cases}
\end{equation}
One can easily check that the gauge-invariant degrees of freedom are then given by \(\mathfrak{I}_1^I, \mathfrak{I}_2^I, \mathfrak{I}_{\alpha}^I, \mathfrak{I}_{\alpha\beta}^I\) and \(\mathfrak{I}_3^i, \mathfrak{I}_4^i, \mathfrak{I}_{\alpha}'^i\), where uppercase indices \(I, J, \ldots\) run from $1$ to $N$, while lowercase indices \(i, j, \ldots\) run from $2$ to $N$. The remaining quantities \(\mathfrak{I}_3^1, \mathfrak{I}_4^1, \mathfrak{I}_{\alpha}'^1\) are pure gauge degrees of freedom and correspond to the two scalars and the transverse vector component of the diffeomorphism vector field \(\xi\).

The choice of the basis transformation~\eqref{eqn:baschange} has another advantage. From assumption~\textit{(\ref{ass:symmetry})} it follows that the field equations must be invariant under arbitrary permutations of the sectors. For the linearized vacuum field equations this implies that the parameter matrices must be invariant under simultaneous permutations of both indices, i.e., under transformations of the form
\begin{equation}\label{eqn:parmattrans}
O^{IJ} \mapsto O^{KL}\sigma^I{}_K\sigma^J{}_L\,,
\end{equation}
for arbitrary permutation matrices \(\sigma\). It then follows that the entries \(O^{IJ}\) must be independent of the individual values of the indices \(I, J\), and they may only depend on whether \(I\) and \(J\) are equal or not. This is the case if and only if the parameter matrices are of the form
\begin{equation}\label{eqn:parmatsym}
O^{IJ} = O^- + (O^+ - O^-)\delta^{IJ}
\end{equation}
with diagonal entries \(O^+\) and off-diagonal entries \(O^-\) for \(O = P, Q, R, M, N\). An explicit calculation shows that the matrix \(\mat{U}\) simultaneously diagonalizes the parameter matrices, so that
\begin{equation}\label{eqn:parmatdiag}
\mat{\mathfrak{O}} = \mat{U} \cdot \mat{O} \cdot \mat{U}^{-1} = \mathrm{diag}(O_1, O_0, \ldots, O_0)\,,
\end{equation}
where \(O_0 = O^+ - O^-\) and \(O_1 = O^+ + (N - 1)O^-\) are the eigenvalues of \(O\). Further introducing \(\vek{\mathfrak{h}}_{ab} = \mat{U} \cdot \vek{h}_{ab}\), the most general linearized field equations can be written in the equivalent form
\begin{equation}\label{eqn:linvaceom2}
0 = \vek{\mathfrak{K}}_{ab} = \mat{\mathfrak{P}} \cdot \partial^p\partial_{(a}\vek{\mathfrak{h}}_{b)p} + \mat{\mathfrak{Q}} \cdot \square\vek{\mathfrak{h}}_{ab} + \mat{\mathfrak{R}} \cdot \partial_a\partial_b\vek{\mathfrak{h}} + \mat{\mathfrak{M}} \cdot \partial^p\partial^q\vek{\mathfrak{h}}_{pq}\eta_{ab} + \mat{\mathfrak{N}} \cdot \square\vek{\mathfrak{h}}\eta_{ab}\,,
\end{equation}
where the parameter matrices \(\mat{\mathfrak{P}}, \mat{\mathfrak{Q}}, \mat{\mathfrak{R}}, \mat{\mathfrak{M}}, \mat{\mathfrak{N}}\) are now diagonal matrices of the form~\eqref{eqn:parmatdiag}. Thus, the equations decouple and we can write them as
\begin{subequations}\label{eqn:linvaceom3}
\begin{align}
0 &= \mathfrak{K}^1_{ab} = P_1\partial^p\partial_{(a}\mathfrak{h}^1_{b)p} + Q_1\square\mathfrak{h}^1_{ab} + R_1\partial_a\partial_b\mathfrak{h}^1 + M_1\partial^p\partial^q\mathfrak{h}^1_{pq}\eta_{ab} + N_1\square\mathfrak{h}^1\eta_{ab}\,,\\
0 &= \mathfrak{K}^i_{ab} = P_0\partial^p\partial_{(a}\mathfrak{h}^i_{b)p} + Q_0\square\mathfrak{h}^i_{ab} + R_0\partial_a\partial_b\mathfrak{h}^i + M_0\partial^p\partial^q\mathfrak{h}^i_{pq}\eta_{ab} + N_0\square\mathfrak{h}^i\eta_{ab}\,.
\end{align}
\end{subequations}
A similar decomposition can be applied to equations~\eqref{eqn:scalar},~\eqref{eqn:vector} and~\eqref{eqn:tensor} in terms of the quantities \(\vek{\mathfrak{I}} = \mat{U} \cdot \vek{I}\). These equations are invariant under gauge transformations if and only if they can be expressed in terms of gauge-invariant quantities only, i.e., they must not depend on the gauge-dependent quantities \(\mathfrak{I}_3^1\), \(\mathfrak{I}_4^1\) and \(\mathfrak{I}_{\alpha}'^1\). This is the case if and only if the eigenvalues of the parameter matrices satisfy the conditions
\begin{equation}\label{eqn:gaugecond}
P_1 + 2Q_1 = P_1 + 2R_1 = M_1 + N_1 = 0\,,
\end{equation}
as we have shown explicitly in~\cite{Hohmann:2009bi}. In the following we will consider only multimetric theories which satisfy these gauge invariance conditions.

\subsection{Bianchi identities}\label{subsec:bianchi}
In the preceding subsection we have shown that the parameter matrices \(\mat{P}, \mat{Q}, \mat{R}, \mat{M}, \mat{N}\) are significantly restricted by the assumptions \textit{(\ref{ass:action})}-\textit{(\ref{ass:flat})} stated in the introduction. In this subsection we will derive further restrictions which originate from the Bianchi identities. These are expected to hold since the theories we consider are derived from an action according to assumption~\textit{(\ref{ass:action})}. Writing the gravitational part \(S_G[g^1, \ldots, g^N]\) of the action as an integral of the Lagrangian density \(\mathcal{L}\) and demanding that it is invariant under diffeomorphisms generated by an arbitrary vector field \(\xi\) leads to the condition
\begin{equation}
0 = \delta_{\xi}S_G = -2\int d^4x\sum_{I = 1}^N\sqrt{g^I}\nabla^I_a\left(\frac{1}{\sqrt{g^I}}\frac{\delta\mathcal{L}}{\delta g^I_{ab}}\right)g^I_{bc}\xi^c\,.
\end{equation}
Using the perturbation ansatz~\eqref{eqn:perturb} and the basis transformation~\eqref{eqn:baschange} this reduces to the linearized Bianchi identity
\begin{equation}\label{eqn:bianchi1}
0 = \partial^a\mathfrak{K}^1_{ab} = \left(\frac{1}{2}P_1 + Q_1\right)\square\partial^a\mathfrak{h}^1_{ab} + \left(\frac{1}{2}P_1 + M_1\right)\partial_b\partial^p\partial^q\mathfrak{h}^1_{pq} + (R_1 + N_1)\square\partial_b\mathfrak{h}^1\,,
\end{equation}
which is a geometric identity. It then follows that the parameters must satisfy the additional constraints
\begin{equation}\label{eqn:bianchicond1}
P_1 = -2Q_1 = -2M_1\,, \qquad R_1 = -N_1\,.
\end{equation}
Together with the gauge invariance conditions~\eqref{eqn:gaugecond} these constraints restrict the eigenvalues \(P_1, Q_1, R_1, M_1, N_1\) of the parameter matrices to only one free parameter, e.g., \(P_1\), and the remaining parameters are determined as
\begin{equation}\label{eqn:bianchigauge}
P_1 = -2Q_1 = -2R_1 = -2M_1 = 2N_1\,.
\end{equation}
Note that in the single-metric case, in which the parameter matrices \(\mat{P}, \mat{Q}, \mat{R}, \mat{M}, \mat{N}\) are replaced by their unique eigenvalues \(P_1, Q_1, R_1, M_1, N_1\), this determines the linearized field equations of a single-metric theory to be identical to those of general relativity, up to a constant factor.

The situation is different for the remaining linear combinations \(\mathfrak{h}^i_{ab}\) of the metric perturbations. For these we do not obtain any constraints from the diffeomorphism-invariance of the gravitational action, since they are gauge invariants according to~\eqref{eqn:gatra}. However, we do obtain constraints from the fact that the matter action \(S_M[g^I, \Psi^I]\) for each of the standard model copies is diffeomorphism invariant and so the corresponding energy-momentum tensors \(T^I_{ab}\) are conserved. Since the field equations are of the form \(\vek{K}_{ab} = 8\pi G_N\vek{T}_{ab}\) according to assumption~\textit{(\ref{ass:tensor})}, this leads to the remaining linearized Bianchi identities
\begin{equation}\label{eqn:bianchi2}
0 = \partial^a\mathfrak{K}^i_{ab} = \left(\frac{1}{2}P_0 + Q_0\right)\square\partial^a\mathfrak{h}^i_{ab} + \left(\frac{1}{2}P_0 + M_0\right)\partial_b\partial^p\partial^q\mathfrak{h}^i_{pq} + (R_0 + N_0)\square\partial_b\mathfrak{h}^i\,,
\end{equation}
which must be satisfied by all solutions of the linearized field equations~\eqref{eqn:linvaceom}. Note that in contrast to the Bianchi identity~\eqref{eqn:bianchi1}, which follows from the diffeomorphism-invariance of the gravitational action, these additional Bianchi identities are in general not geometric identities, and are not necessarily satisfied by arbitrary metric perturbations. In order to satisfy~\eqref{eqn:bianchi2} we are thus left with two possibilities:
\begin{enumerate}[(i)]
\item
\textit{The Bianchi identities~\eqref{eqn:bianchi2} are geometric identities and satisfied by arbitrary perturbations \(\mathfrak{h}^i_{ab}\) of the metric tensors.}\\
This is the case if and only if the eigenvalues of the parameter matrices satisfy the additional constraints
\begin{equation}\label{eqn:bianchicond2}
P_0 = -2Q_0 = -2M_0\,, \qquad R_0 = -N_0\,.
\end{equation}
It then follows that we are left with only three free parameters, e.g., \(P_1, P_0, R_0\).

\item
\textit{The Bianchi identities~\eqref{eqn:bianchi2} are not geometric identities and satisfied only by solutions of the gravitational field equations.}\\
In this case we do not obtain any additional constraints on the parameter matrices. Note that even in this less restrictive case the Bianchi identities~\eqref{eqn:bianchi2} are implied by the gravitational field equations, taking into account that the source of the gravitational field is given by the matter energy momentum tensors, which must be divergence-free due to the diffeomorphism-invariance of the matter action. Thus,~\eqref{eqn:bianchi2} must be satisfied by all solutions of the gravitational field equations.
\end{enumerate}
In the remainder of this article we will not assume that the Bianchi conditions~\eqref{eqn:bianchicond2} on the parameters are satisfied in general, and instead use the smaller set~\eqref{eqn:bianchigauge} of combined Bianchi and gauge conditions. We will explicitly show how the different sets of conditions influence the possible wave-like solutions in the following section.

\subsection{Wave ansatz}\label{subsec:wave}
We are now in the position to explicitly construct wave-like solutions to the gauge invariant field equations~\eqref{eqn:scalar},~\eqref{eqn:vector} and~\eqref{eqn:tensor}. For simplicity, we apply the basis transformation~\eqref{eqn:baschange}, so that all occurring parameter matrices become diagonal and the field equations are written in terms of the gauge-invariant quantities \(\vek{\mathfrak{I}}\). For these quantities we use the wave ansatz
\begin{equation}\label{eqn:wave}
\vek{\mathfrak{I}} = \vek{\hat{\mathfrak{I}}}\exp(ik_ax^a)\,,
\end{equation}
for a single Fourier mode, where \(\vek{\hat{\mathfrak{I}}}\) are constants which determine the amplitude of the wave and \(k_a\) is a constant covector. The aim of this section is to compute for which amplitudes and wave covectors the field equations are satisfied.

First we solve the tensor equations~\eqref{eqn:tensor}. Using the basis transformation~\eqref{eqn:baschange}, these take the form
\begin{equation}
0 = 2Q_1\square\mathfrak{I}^1_{\alpha\beta}\,, \quad 0 = 2Q_0\square\mathfrak{I}^i_{\alpha\beta}\,.
\end{equation}
If one of the eigenvalues \(Q_0, Q_1\) vanishes, the corresponding equation is satisfied identically. In this case the linearized field equations~\eqref{eqn:linvaceom} are not sufficient to solve for some of the modes \(\mathfrak{I}^I_{\alpha\beta}\) and a calculation based on the full nonlinear field equations is required. We will not attempt such a calculation in this article and restrict ourselves to the case that both \(Q_0\) and \(Q_1\) are nonzero. We then insert the wave ansatz~\eqref{eqn:wave} and obtain the equations
\begin{equation}
0 = -2Q_1k_ak^a\mathfrak{I}^1_{\alpha\beta}\,, \quad 0 = -2Q_0k_ak^a\mathfrak{I}^i_{\alpha\beta}\,.
\end{equation}
We immediately see that non-vanishing wave-like solutions exist if and only if the wave covector \(k_a\) is null, \(k_ak^a = 0\), while the amplitudes \(\mathfrak{I}^I_{\alpha\beta}\) may be arbitrary.

Second we consider the vector equations~\eqref{eqn:vector}. After applying the basis transformation and inserting the wave ansatz, these can be written in the form
\begin{subequations}
\begin{align}
0 &= \left(\begin{array}{cc}
\frac{1}{2}P_1k_{\alpha}k^{\alpha} & 0\\
-\frac{i}{2}P_1k_0 & 0
\end{array}\right) \cdot \left(\begin{array}{c}
\mathfrak{I}^1_{\alpha}\\
\mathfrak{I}'^1_{\alpha}
\end{array}\right),\label{eqn:vector1}\\
0 &= \left(\begin{array}{cc}
\frac{1}{2}P_0k_0^2 - Q_0k_ak^a & -\frac{i}{2}(P_0 + 2Q_0)k_0k_ak^a\\
-\frac{i}{2}P_0k_0 & -\frac{1}{2}(P_0 + 2Q_0)k_ak^a
\end{array}\right) \cdot \left(\begin{array}{c}
\mathfrak{I}^i_{\alpha}\\
\mathfrak{I}'^i_{\alpha}
\end{array}\right),\label{eqn:vector0}
\end{align}
\end{subequations}
where we already used the gauge invariance conditions~\eqref{eqn:gaugecond} to eliminate \(Q_1\). Consequently, the pure gauge quantity \(\mathfrak{I}'^1_{\alpha}\) drops out and the field equations depend only on the physical degrees of freedom. From equation~\eqref{eqn:vector1} we see that we can solve for the quantities \(\mathfrak{I}^1_{\alpha}\) using the linearized field equations only if \(P_1 \neq 0\). It then follows that there are no wave-like solutions for \(\mathfrak{I}^1_{\alpha}\), since the field equations require that both \(k_0\) and \(k_{\alpha}\) must vanish. Similarly, a solvable equation for \(\mathfrak{I}^i_{\alpha}\) and \(\mathfrak{I}'^i_{\alpha}\) requires that both \(Q_0\) and \(P_0 + 2Q_0\) are nonzero. It then follows that wave-like solutions exist if and only if the determinant of the matrix in equation~\eqref{eqn:vector0} vanishes,
\begin{equation}
\frac{1}{2}Q_0(P_0 + 2Q_0)(k_ak^a)^2 = 0\,,
\end{equation}
i.e., for null waves, \(k_ak^a = 0\). For these solutions \(\mathfrak{I}'^i_{\alpha}\) is always allowed to be nonzero, while \(\mathfrak{I}^i_{\alpha}\) is allowed to be nonzero only if \(P_0 = 0\). Note that \(P_0 + 2Q_0\) for theories in which the Bianchi identities~\eqref{eqn:bianchi2} are geometric identities and in which the parameters satisfy the conditions~\eqref{eqn:bianchicond2}. In this case the linearized field equations are not solvable for the vector potential \(\mathfrak{I}'^i_{\alpha}\). The remaining vector potential \(\mathfrak{I}^i_{\alpha}\) must vanish for \(P_0 \neq 0\) and cannot be determined from the linearized equations for \(P_0 = 0\).

Finally we discuss the scalar equations~\eqref{eqn:scalar}. We can proceed in complete analogy to the vector equations shown above. We apply the basis transformation~\eqref{eqn:baschange}, insert the wave ansatz~\eqref{eqn:wave} and make use of the gauge invariance conditions~\eqref{eqn:gaugecond} in order to eliminate the parameters \(Q_1\), \(R_1\) and \(N_1\). The equations for the first component \(\mathfrak{I}^1\) can then be written in matrix form as
\begin{equation}\label{eqn:scalar1}
0 = \left(\begin{array}{cccc}
-(P_1 + 2M_1)k_{\alpha}k^{\alpha} & 4M_1k_{\alpha}k^{\alpha} - 3(P_1 + 2M_1)k_0^2 & 0 & 0\\
0 & 2iP_1k_0 & 0 & 0\\
(P_1 + 6M_1)k_{\alpha}k^{\alpha} & -4(P_1 + 3M_1)k_{\alpha}k^{\alpha} + 3(P_1 + 6M_1)k_0^2 & 0 & 0\\
-P_1 & P_1 & 0 & 0
\end{array}\right) \cdot \left(\begin{array}{c}
\mathfrak{I}^1_1\\
\mathfrak{I}^1_2\\
\mathfrak{I}^1_3\\
\mathfrak{I}^1_4
\end{array}\right).
\end{equation}
We immediately see that these do not depend on the pure gauge quantities \(\mathfrak{I}^1_3\) and \(\mathfrak{I}^1_4\) as a consequence of the gauge invariance conditions. For the gauge-invariant quantities \(\mathfrak{I}^1_1\) and \(\mathfrak{I}^1_2\) we must distinguish two different cases. If \(P_1 \neq 0\), it follows from the second component equation of~\eqref{eqn:scalar1} that there are no wave-like solutions for \(\mathfrak{I}^1_2\). From the last component equation we further see that \(\mathfrak{I}^1_1 = \mathfrak{I}^1_2\) and thus there are no wave-like solutions for \(\mathfrak{I}^1_1\) either. In the case \(P_1 = 0\) the equations are not sufficient to determine the quantities \(\mathfrak{I}^1_1\) and \(\mathfrak{I}^1_2\).

For the remaining quantities \(\mathfrak{I}^i\) we proceed similarly and write the field equations in matrix form. As we already have done for the vector equation~\eqref{eqn:vector0}, we calculate the determinant of the occurring matrix, which takes the form
\begin{equation}\label{eqn:scaldet}
\Big\{3Q_0(P_0 + 2Q_0)\big[Q_0(P_0 + Q_0 + R_0) + M_0(Q_0 - 3R_0) + N_0(3P_0 + 4Q_0)\big]\Big\}(k_ak^a)^4 = 0\,.
\end{equation}
Again we distinguish two cases. If the constant factor in curly brackets vanishes, the linearized field equations are not sufficient to solve for the quantities \(\mathfrak{I}^i\). Otherwise, the field equations can be solved by the wave ansatz if and only if the determinant vanishes, which is the case for null waves. A quick calculation shows that the solutions take the form
\begin{equation}\label{eqn:scalsol}
\mathfrak{I}^i_1 = -\mathfrak{I}^i_2 = \frac{P_0 + 2R_0}{4P_0 + 16R_0}(i\mathfrak{I}^i_3 + 2\mathfrak{I}^i_4)\,.
\end{equation}
Note further that if the parameters satisfy the Bianchi conditions~\eqref{eqn:bianchicond2}, the determinant~\eqref{eqn:scaldet} vanishes identically so that the linearized field equations are not solvable for the scalar potentials~\(\mathfrak{I}^i\).

This result completes our discussion of gravitational waves in the gauge-invariant formalism. We have shown that we need to impose several conditions on the eigenvalues of the parameter matrices \(\mat{P}, \mat{Q}, \mat{R}, \mat{M}, \mat{N}\). From assumption~\textit{(\ref{ass:symmetry})} we concluded that they must be of the form~\eqref{eqn:parmatsym} and can be diagonalized according to equation~\eqref{eqn:parmatdiag}. The gauge conditions~\eqref{eqn:gaugecond} guarantee that the linearized field equations are gauge-invariant and thus depend only on the physical degrees of freedom. The solvability conditions
\begin{gather}
Q_0 \neq 0\,, \quad P_1 = -2Q_1 \neq 0\,, \quad P_0 + 2Q_0 \neq 0\,,\nonumber\\
Q_0(P_0 + Q_0 + R_0) + M_0(Q_0 - 3R_0) + N_0(3P_0 + 4Q_0) \neq 0\label{eqn:solvability}
\end{gather}
allow a complete treatment of gravitational waves using the linearized field equations. Under these conditions, wave-like solutions for the scalar, vector and tensor components of the gauge-invariant quantities exist if and only if the wave covector \(k_a\) is null, \(k_ak^a = 0\). Since we treat gravitational waves as a small perturbation of the flat Minkowski background, whose null directions govern the propagation of light within our approximation, it then follows that the speed of gravitational waves equals the speed of light, \(v_g = c\). In the more general case that a concrete theory does not satisfy all of the solvability conditions~\eqref{eqn:solvability}, a calculation based on the full non-linear field equations is necessary to determine whether additional wave-like solutions exist. Since we do not perform such a calculation in this article, we restrict ourselves to the null wave solutions we have found so far and determine their possible polarizations in the following section.

\section{Newman-Penrose formalism and possible polarizations}\label{sec:newpen}
In the preceding section we have shown that the most general linearized vacuum field equations~\eqref{eqn:linvaceom} can be solved by the wave ansatz~\eqref{eqn:wave} only if the wave covector is null. We will now turn our focus to the possible polarizations of gravitational waves. Since we are dealing only with null waves, the polarizations can easily be decomposed by employing the Newman-Penrose formalism introduced in~\cite{Newman:1961qr}. We will then employ the classification scheme detailed in~\cite{Eardley:1973br,Eardley:1974nw} in order to determine the E(2) class of multimetric gravity, which could be measured by the upcoming gravitational wave experiments. The connection between these experiments and the class of multimetric theories we consider in this article is established by assumption~\textit{(\ref{ass:action})} stated in the introduction. It follows from this assumption that a gravitational wave experiment built up from visible matter, i.e., only one standard model copy \(\Psi^1\), is sensitive only to the corresponding metric \(g^1_{ab}\). Thus, it is sufficient to determine the possible polarizations of wave-like solutions for the metric \(g^1_{ab}\). From assumption~\textit{(\ref{ass:symmetry})} it follows further that the possible polarizations are the same for all sectors~\((\Psi^I, g^I)\).

Basic ingredient of the Newman-Penrose formalism is a convenient double null basis of the tangent space. In the following, we will use the notation of~\cite{Will:1993ns} and denote the basis vectors by \(l^a, n^a, m^a, \bar{m}^a\). In the \(x^0, x^{\alpha}\) basis they take the form
\begin{equation}
l^a = (1, 0, 0, 1)\,, \qquad n^a = \frac{1}{2}(1, 0, 0, -1)\,, \qquad m^a = \frac{1}{\sqrt{2}}(0, 1, i, 0)\,, \qquad \bar{m}^a = \frac{1}{\sqrt{2}}(0, 1, -i, 0)\,.
\end{equation}
In the new basis the flat Minkowski metric takes the form
\begin{equation}
\eta_{ab} = \left(\begin{array}{cccc}
0 & -1 & 0 & 0\\
-1 & 0 & 0 & 0\\
0 & 0 & 0 & 1\\
0 & 0 & 1 & 0
\end{array}\right).
\end{equation}
We now consider a plane wave propagating in the positive \(x^3\) direction. The wave covector then takes the form \(k_a = -\omega l_a\) and the metric perturbations are given by
\begin{equation}\label{eqn:zwave}
\vek{h}_{ab} = \vek{\hat{h}}_{ab}e^{i\omega(x^0 - x^3)} = \vek{\hat{h}}_{ab}e^{i\omega u}
\end{equation}
for the retarded time \(u = x^0 - x^3\). The effect of this wave on a set of test masses consisted by one type \(\Psi^1\) of standard model matter depends only on the Riemann tensor of the corresponding metric \(g^1_{ab}\). As shown in~\cite{Eardley:1974nw}, the Riemann tensor of a plane wave is determined completely by the six so-called electric components. For the wave~\eqref{eqn:zwave}, these can be written as
\begin{gather}
\vek{\Psi}_2 = -\frac{1}{6}\vek{R}_{nlnl} = \frac{1}{12}\vek{\ddot{h}}_{ll}\,, \quad \vek{\Psi}_3 = -\frac{1}{2}\vek{R}_{nln\bar{m}} = -\frac{1}{2}\overline{\vek{R}_{nlnm}} = \frac{1}{4}\vek{\ddot{h}}_{l\bar{m}} = \frac{1}{4}\overline{\vek{\ddot{h}}_{lm}}\,,\nonumber\\
\vek{\Psi}_4 = -\vek{R}_{n\bar{m}n\bar{m}} = -\overline{\vek{R}_{nmnm}} = \frac{1}{2}\vek{\ddot{h}}_{\bar{m}\bar{m}} = \frac{1}{2}\overline{\vek{\ddot{h}}_{mm}}\,, \quad \vek{\Phi}_{22} = -\vek{R}_{nmn\bar{m}} = \frac{1}{2}\vek{\ddot{h}}_{m\bar{m}}\,,\label{eqn:riemcomp}
\end{gather}
where dots denote derivatives with respect to \(u\). We now examine which of the components~\eqref{eqn:riemcomp} may occur for gravitational waves satisfying the linearized field equations~\eqref{eqn:linvaceom}. Inserting the wave ansatz~\eqref{eqn:zwave} we immediately see that the terms containing \(\mat{Q}\) and \(\mat{N}\) drop out, since \(\square\vek{h}_{ab} = 0\) for a null wave. Writing the curvature tensor \(\vek{K}_{ab}\) in the Newman-Penrose basis, we find that the five component equations
\begin{equation}
0 = \vek{K}_{ll} = \vek{K}_{mm} = \vek{K}_{\bar{m}\bar{m}} = \vek{K}_{lm} = \vek{K}_{l\bar{m}}
\end{equation}
are satisfied identically, while the remaining five component equations take the form
\begin{subequations}\label{eqn:np_wave}
\begin{align}
0 &= \vek{K}_{nn} = 2\mat{R} \cdot \ddot{\vek{h}}_{m\bar{m}} - (\mat{P} + 2\mat{R}) \cdot \ddot{\vek{h}}_{ln}\,,\label{eqn:np_nn}\\
0 &= \vek{K}_{ln} = -\frac{1}{2}(\mat{P} + 2\mat{M}) \cdot \ddot{\vek{h}}_{ll}\,,\label{eqn:np_ln}\\
0 &= \vek{K}_{nm} = -\frac{1}{2}\mat{P} \cdot \ddot{\vek{h}}_{lm}\,,\label{eqn:np_nm}\\
0 &= \vek{K}_{n\bar{m}} = -\frac{1}{2}\mat{P} \cdot \ddot{\vek{h}}_{l\bar{m}}\,,\label{eqn:np_nbarm}\\
0 &= \vek{K}_{m\bar{m}} = \mat{M} \cdot \ddot{\vek{h}}_{ll}\label{eqn:np_mbarm}\,.
\end{align}
\end{subequations}
Recall that the parameter matrices can be brought into diagonal form using the basis transformation~\eqref{eqn:parmatdiag} so that the field equations decouple as shown in~\eqref{eqn:linvaceom3}. Applying this decomposition to equation~\eqref{eqn:np_nn} we obtain
\begin{subequations}
\begin{align}
0 &= \mathfrak{K}^1_{nn} = 2R_1 \cdot \ddot{\mathfrak{h}}^1_{m\bar{m}} - (P_1 + 2R_1) \cdot \ddot{\mathfrak{h}}^1_{ln}\,,\\
0 &= \mathfrak{K}^i_{nn} = 2R_0 \cdot \ddot{\mathfrak{h}}^i_{m\bar{m}} - (P_0 + 2R_0) \cdot \ddot{\mathfrak{h}}^i_{ln}\,,
\end{align}
\end{subequations}
and similarly for the remaining four component equations. Let \(h_{ab}\) denote one of the metric perturbations \(\mathfrak{h}^I_{ab}\) and \(P, R, M\) the corresponding eigenvalues of the parameter matrices \(\mat{P}, \mat{R}, \mat{M}\). We distinguish the following cases:
\begin{itemize}
\item
\(M = P = 0\): In this case equations~\eqref{eqn:np_ln} and~\eqref{eqn:np_mbarm} are satisfied identically for arbitrary amplitudes \(\hat{h}_{ll}\). For waves of this type the corresponding component \(R_{nlnl} = -6\Psi_2\) of the Riemann tensor is allowed to be nonzero. Following the classification detailed in~\cite{Eardley:1974nw}, they belong to the E(2) class \(\mathrm{II}_6\).

\item
\(M \neq 0\) and \(P = 0\): Equation~\eqref{eqn:np_mbarm} forbids waves with a nonzero amplitude \(\hat{h}_{ll}\), and thus \(\Psi_2 = 0\). Equations~\eqref{eqn:np_nm} and~\eqref{eqn:np_nbarm} are satisfied identically for arbitrary amplitudes \(\hat{h}_{lm}\) and \(\hat{h}_{l\bar{m}}\). It then follows that \(R_{nln\bar{m}} = -2\Psi_3\) is allowed to be nonzero. Waves of this type belong to the E(2) class \(\mathrm{III}_5\).

\item
\(P \neq 0\) and \(P + 2R \neq 0\): For \(P \neq 0\) it follows from equations~\eqref{eqn:np_ln}, \eqref{eqn:np_nm}, \eqref{eqn:np_nbarm} and~\eqref{eqn:np_mbarm} that \(\hat{h}_{ll} = \hat{h}_{lm} = \hat{h}_{l\bar{m}} = 0\), and thus \(\Psi_2 = \Psi_3 = 0\). The remaining equation~\eqref{eqn:np_nn} is solved for \(2R\hat{h}_{m\bar{m}} = (P + 2R)\hat{h}_{ln}\). Hence, the corresponding component \(R_{nmn\bar{m}} = -\Phi_{22}\) of the Riemann tensor is allowed to be nonzero and the wave belongs to E(2) class \(\mathrm{N}_3\).

\item
\(P = -2R \neq 0\): This is the most restrictive case. Equations~\eqref{eqn:np_wave} are satisfied only for \(\hat{h}_{ll} = \hat{h}_{lm} = \hat{h}_{l\bar{m}} = \hat{h}_{m\bar{m}} = 0\), and thus \(\Psi_2 = \Psi_3 = \Phi_{22} = 0\). The only allowed polarization is \(R_{n\bar{m}n\bar{m}} = -\Psi_4\) and the wave belongs to E(2) class \(\mathrm{N}_2\).
\end{itemize}
The classification can be summarized in a convenient graphical form. The following diagram shows how the E(2) class is determined by the values of the parameters \(P, R, M\) in the linearized field equations:
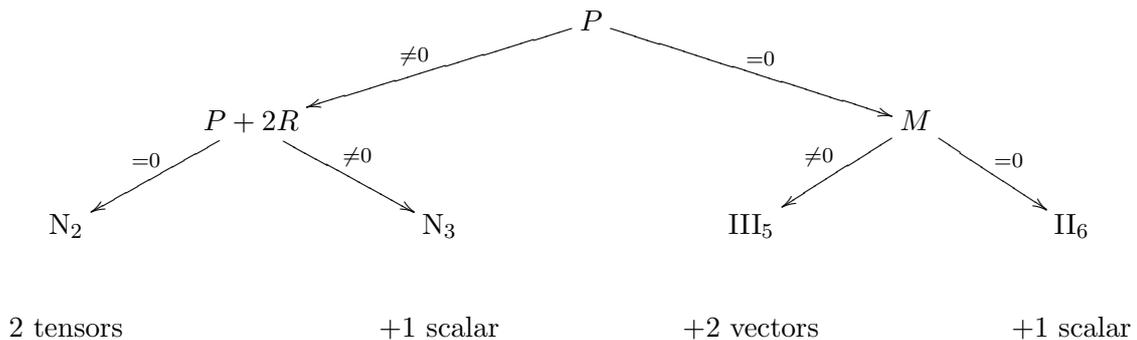
\begin{equation*}
\xymatrix{&&& P \ar[dll]_{\neq 0} \ar[drr]^{= 0} &&&\\
& P + 2R \ar[dl]_{= 0} \ar[dr]^{\neq 0} &&&& M \ar[dl]_{\neq 0} \ar[dr]^{= 0} &\\
\mathrm{N}_2 && \mathrm{N}_3 && \mathrm{III}_5 && \mathrm{II}_6\\
\text{2 tensors} && \text{+1 scalar} && \text{+2 vectors} && \text{+1 scalar}}
\end{equation*}
The parameters \(P, R, M\) in this diagram are either the set \(P_1, R_1, M_1\), which yields the E(2) class for waves of type \(\mathfrak{h}^1_{ab}\), or \(P_0, R_0, M_0\), which instead yields the E(2) class for waves of type \(\mathfrak{h}^i_{ab}\). We thus obtain two E(2) classes for the different linear combinations of the metric perturbations \(\vek{h}_{ab}\).

We finally turn our focus to the viewpoint of a physical observer. From assumption~\textit{(\ref{ass:action})} in the introduction it follows that any experimental setup consisting of visible matter only, i.e., of only one copy \(\Psi^1\) of the standard model, is affected by only one metric tensor \(g^1_{ab}\). A gravitational wave detector consisting of visible matter can thus measure only one of the Riemann tensors \(R^1_{abcd}\). From the basis transformation~\eqref{eqn:baschange} it follows further that each Riemann tensor \(R^I_{abcd}\) depends on all metric perturbations \(\mathfrak{h}^I_{ab}\). As a consequence, it is not possible to measure the metric perturbations \(\mathfrak{h}^I_{ab}\) separately. A gravitational wave experiment can only indicate whether a wave with certain polarization exists for any of the linear combinations \(\mathfrak{h}^I_{ab}\). Thus, only the larger of the two aforementioned E(2) classes can be determined.

\section{Examples}\label{sec:examples}
In the previous section~\ref{sec:newpen} we constructed a formalism to calculate the E(2) class of multimetric gravity theories that determines the possible polarizations of gravitational waves. It turned out that the E(2) class is fully determined by the eigenvalues of the parameter matrices \(\mat{P}, \mat{Q}, \mat{R}, \mat{M}, \mat{N}\) in the linearized vacuum field equations~\eqref{eqn:linvaceom}. We will now apply this classification to a number of example theories.

\subsection{General relativity}
Although general relativity is not a multimetric theory, we can apply a slightly modified version of the calculations presented in this article. For the case of \(N = 1\) metric tensors, we replace the parameter matrices in the linearized vacuum field equations by their unique eigenvalues. For general relativity, these take the values
\begin{equation}\label{eqn:grparams}
P_1 = 1\,, \quad Q_1 = R_1 = M_1 = -\frac{1}{2}\,, \quad N_1 = \frac{1}{2}\,.
\end{equation}
One easily checks that they satisfy the gauge invariance conditions~\eqref{eqn:gaugecond} and the Bianchi conditions~\eqref{eqn:bianchicond1}. Following the calculation presented in subsection~\ref{subsec:wave}, one finds that the wave solutions are completely determined by the linearized field equations, and that the only permitted non-zero amplitude is the tensor \(\mathfrak{I}^1_{\alpha\beta}\). Finally, a comparison of the parameters~\eqref{eqn:grparams} with the diagram at the end of section~\ref{sec:newpen} correctly shows that the E(2) class of general relativity is \(\mathrm{N}_2\). Note that this is the generic case for single-metric theories due to the parameter constraints~\eqref{eqn:bianchigauge}.

\subsection{A simple multimetric theory}
A simple class of multimetric gravity theories with \(N \geq 2\) metrics, which also contains the theories presented in~\cite{Hohmann:2010vt,Hohmann:2010ni}, is given by the gravitational action
\begin{multline}\label{eqn:gravaction}
S_G[g^1, \ldots, g^N] = \frac{1}{16\pi}\int d^4x\sqrt{g_0}\Bigg[x\sum_{I, J = 1}^{N}g^{I\,ij}R^J{}_{ij} + \sum_{I = 1}^Ng^{I\,ij}\Bigg(yR^I{}_{ij}\\
+ u\tilde{S}^I{}_i\tilde{S}^I{}_j + v\tilde{S}^I{}_k\tilde{S}^{I\,k}{}_{ij} + w\tilde{S}^{I\,k}{}_{im}\tilde{S}^{I\,m}{}_{jk} + g^{I\,kl}g^I{}_{mn}\Big(r\tilde{S}^{I\,m}{}_{ik}\tilde{S}^{I\,n}{}_{jl} + s\tilde{S}^{I\,m}{}_{ij}\tilde{S}^{I\,n}{}_{kl}\Big)\Bigg)\Bigg],
\end{multline}
where the connection difference tensors \(\tilde{S}^{I\,k}{}_{ij}\), \(\tilde{S}^{I}{}_i\) are defined as
\begin{equation}
S^{IJ\,i}{}_{jk} = \Gamma^{I\,i}{}_{jk} - \Gamma^{J\,i}{}_{jk}\,,\quad
S^{IJ}{}_j = S^{IJ\,k}{}_{jk}\,,\quad
\tilde S^{J\,i}{}_{jk} = \frac{1}{N}\sum_{I = 1}^N S^{IJ\,i}{}_{jk}\,,\quad
\tilde S^{J}{}_j = \tilde S^{J\,k}{}_{jk}\,,
\end{equation}
the volume form is given by \(g_0 = \prod_{I = 1}^{N}\left(g^I\right)^{1/N}\), and \(x, y, u, v, w, r, s\) are constant parameters. Starting from the action~\eqref{eqn:gravaction}, we derive the gravitational field equations by variation with respect to the metric tensors and use the perturbation ansatz~\eqref{eqn:perturb} to keep only the terms of linear order~\(\mathcal{O}(h)\). This yields the eigenvalues of the parameter matrices
\begin{gather}
P_1 = -2Q_1 = -2R_1 = -2M_1 = 2N_1 = Nx + y\,, \quad R_0 = M_0 = \frac{Nx - v + 2s}{2}\,,\nonumber\\
P_0 = -Nx + y - w + r - 2s\,, \quad Q_0 = \frac{Nx - y + w - 3r}{2}\,, \quad N_0 = \frac{-Nx - y - u + v - s}{2}\,,\label{eqn:multiparams1}
\end{gather}
which satisfy the gauge invariance conditions~\eqref{eqn:gaugecond} and the Bianchi conditions~\eqref{eqn:bianchicond1} since we started from a diffeomorphism-invariant action. The extended Bianchi conditions~\eqref{eqn:bianchicond2} are satisfied if and only if the parameters satisfy the constraints
\begin{equation}\label{eqn:biparamcond}
0 = s + r = u + v + w = u + y - s\,.
\end{equation}
It thus follows from our discussion of the Bianchi identities in subsection~\ref{subsec:bianchi} that in the generic case, in which the conditions~\eqref{eqn:biparamcond} are not satisfied, the Bianchi identities~\eqref{eqn:bianchi2} are not geometric identities, but satisfied only by solutions of the gravitational field equations. Next, we apply the linearized multimetric extension of the parametrized post-Newtonian formalism detailed in~\cite{Hohmann:2010ni}. Consistency with solar system measurements of the PPN parameters requires
\begin{equation}
y = \frac{1}{2 - N} - Nx\,, \quad v = \frac{6 - N}{4 - 2N} - Nx + 2u\,, \quad w = -\frac{6 - N}{4 - 2N} + Nx - 3u\,, \quad r = -\frac{1}{2 - N} + Nx - u\,,
\end{equation}
which leaves us with only three free parameters \(x, u, s\) and restricts the number of metrics to \(N > 2\). In terms of these remaining parameters the eigenvalues of the parameter matrices take the values
\begin{gather}
P_1 = -2Q_1 = -2R_1 = -2M_1 = 2N_1 = \frac{1}{2 - N}\,, \quad R_0 = M_0 = -\frac{6 - N}{8 - 4N} + Nx - u + s\,,\nonumber\\
P_0 = \frac{6 - N}{4 - 2N} - 2Nx + 2u - 2s\,, \quad Q_0 = -\frac{1}{4}\,, \quad N_0 = \frac{4 - N}{8 - 4N} + \frac{-Nx + u - s}{2}\,.\label{eqn:multiparams2}
\end{gather}
Using the calculation of subsection~\ref{subsec:wave}, we find the following wave solutions for the gauge invariant potentials \(\mathfrak{I}\):
\begin{itemize}
\item \textit{Tensor modes:}\\
Since both \(Q_1\) and \(Q_0\) are nonzero, wave-like solutions for all tensor potentials \(\mathfrak{I}^I_{\alpha\beta}\) exist.
\item \textit{Vector modes:}\\
From \(P_1 \neq 0\) it follows that there are no wave-like solutions for the vector potential \(\mathfrak{I}^1_{\alpha}\). If the parameters satisfy the condition
\begin{equation}\label{eqn:paramcond1}
Nx - u + s = \frac{1}{2 - N}\,,
\end{equation}
the linearized field equations are not sufficient to determine the vector potentials \(\mathfrak{I}'^i_{\alpha}\) and the remaining vector potentials \(\mathfrak{I}^i_{\alpha}\) must vanish. Otherwise, wave-like solutions for the vector potentials \(\mathfrak{I}'^i_{\alpha}\) exist, and wave-like solutions for \(\mathfrak{I}^i_{\alpha}\) exist if and only if the parameters satisfy
\begin{equation}\label{eqn:paramcond2}
Nx - u + s = \frac{6 - N}{8 - 4N}\,.
\end{equation}
\item \textit{Scalar modes:}\\
From \(P_1 \neq 0\) it follows that there are no wave-like solutions for the scalar potentials \(\mathfrak{I}^1_1, \mathfrak{I}^1_2\). If the parameters satisfy the conditions~\eqref{eqn:paramcond1}, the linearized field equations are not sufficient to determine the scalar potentials \(\mathfrak{I}^i_1, \mathfrak{I}^i_2, \mathfrak{I}^i_3, \mathfrak{I}^i_4\). Otherwise, wave-like solutions for \(\mathfrak{I}^i_1, \mathfrak{I}^i_2, \mathfrak{I}^i_3, \mathfrak{I}^i_4\) exist and satisfy~\eqref{eqn:scalsol}.
\end{itemize}
Finally, we determine the E(2) class of our example theory. A comparison of the eigenvalues~\eqref{eqn:multiparams2} with the classification detailed in section~\ref{sec:newpen} shows that the E(2) class for the symmetric linear combinations \(\mathfrak{h}^1_{ab}\) of metric perturbations, for which the eigenvalues \(P_1, R_1, M_1\) of the parameter matrices are relevant, is \(\mathrm{N}_2\). The E(2) class for the remaining linear combinations \(\mathfrak{h}^i_{ab}\), and thus the effective E(2) class of the theory, is \(\mathrm{II}_6\) if the parameters satisfy the condition~\eqref{eqn:paramcond2}, and \(\mathrm{N}_2\) otherwise. This means that the only polarizations that can be measured by a gravitational wave experiment are the two tensor polarizations which are also present in general relativity, unless the parameters satisfy~\eqref{eqn:paramcond2}, in which case all six possible polarizations of gravitational waves may be present.

\section{Conclusion}\label{sec:conclusion}
In this article we have discussed the propagation of gravitational waves in theories with \({N \geq 2}\) metric tensors and a corresponding number of standard model copies. These theories were designed to explain the observed cosmological late-time acceleration while being consistent with solar system experiments. We have examined two characteristic properties of gravitational waves: their propagation velocity and their polarization. We have shown that in a weak-field approximation around a flat, maximally symmetric Minkowski background all gravitational waves propagate at the speed of light. Using the Newman-Penrose formalism we found that there are always two tensor modes; in addition two vector modes and two scalar modes may exist. In terms of E(2) representations this means that multimetric gravity theories can be of class \(\mathrm{N}_2\), \(\mathrm{N}_3\), \(\mathrm{III}_5\) or \(\mathrm{II}_6\).

We then applied our construction to two examples. First, we discussed general relativity and showed that our formalism can also be applied to the special case \(N = 1\) of a single-metric gravity theory. We re-obtained the well-known result that general relativity is of class \(\mathrm{N}_2\), i.e., there are only two tensor polarizations of gravitational waves. Second, we applied our construction to a class of multimetric gravity theories including the theories proposed in~\cite{Hohmann:2010vt,Hohmann:2010ni} and showed that these are either of class \(\mathrm{N}_2\) or \(\mathrm{II}_6\), depending on the choice of parameters.

Our results connect the theoretical framework of multimetric gravity to the physics of gravitational waves, which is the subject of several current and upcoming experiments. It is expected that these will be able to measure both the propagation velocity and the polarization of gravitational waves, which are the two properties addressed in this article. A question of particular interest is whether scalar or vector polarizations will be detected. While these do not exist in general relativity, they are allowed in certain multimetric gravity theories.

Now that we examined the propagation of gravitational waves in multimetric gravity theories, the next task that should be performed is to discuss their production by sources such as binary stars. Further research on this topic should show which of the propagating wave polarizations are emitted from a given source. Moreover, quantitative calculations should yield the amplitude of the emitted waves in the different metric sectors. Since energy may be emitted in all metric sectors, but only one of them is visible to gravitational wave detectors, one might expect a difference between the directly observed energy emission and the total energy loss inferred from the orbital decay. This could provide another testbed for multimetric gravity both by the upcoming gravitational wave experiments~\cite{Thorne:1987af,Sathyaprakash:2009xs} and existing observations of binary pulsars~\cite{Hulse:1974eb,Weisberg:2010,Lorimer:2008se}.

Finally, it should be examined how the presence of cosmic gravitational fields affects both electromagnetic and gravitational radiation as they propagate from a common source, such as a supernova or a binary pulsar, towards our solar system. Calculations of this type are crucial for the interpretation of experiments which compare the arrival times of both types of radiation, as they might undergo a different Shapiro delay~\cite{Desai:2008vj,Kahya:2010dk} or a different gravitational lensing.

\acknowledgments
The author is happy to thank Shantanu Desai, Matthias Lange, Christian Pfeifer, Luigi Pilo, Gunnar Prei\ss{}, Joel Weisberg and Mattias Wohlfarth for helpful comments. He gratefully acknowledges partial financial support from the German Research Foundation DFG through the Emmy Noether grant WO 1447/1-1.

%%%%%%%%%%%%%%%%%%%%%%%%%%%%%%%%%%%%%%%%%%%


\begin{thebibliography}{00}
%\cite{Hohmann:2009bi}
\bibitem{Hohmann:2009bi}
  M.~Hohmann and M.~N.~R.~Wohlfarth,
  %``No-go theorem for bimetric gravity with positive and negative mass,''
  Phys.\ Rev.\ D {\bf 80} (2009) 104011,
  [arXiv:0908.3384 [gr-qc]].

%\cite{Hohmann:2010vt}
\bibitem{Hohmann:2010vt}
  M.~Hohmann and M.~N.~R.~Wohlfarth,
  %``Repulsive gravity model for dark energy,''
  Phys.\ Rev.\ D {\bf 81} (2010) 104006,
  [arXiv:1003.1379 [gr-qc]].

%\cite{Hohmann:2010ni}
\bibitem{Hohmann:2010ni}
  M.~Hohmann and M.~N.~R.~Wohlfarth,
  %``Multimetric extension of the PPN formalism: experimental consistency of repulsive gravity,''
  Phys.\ Rev.\ D\ {\bf 82} (2010) 084028,
  [arXiv:1007.4945 [gr-qc]].

%\cite{arXiv:0912.0790}
\bibitem{arXiv:0912.0790}
  M.~Milgrom,
  %``Bimetric MOND gravity,''
  Phys.\ Rev.\ D\ {\bf 80} (2009) 123536,
  [arXiv:0912.0790 [gr-qc]].
  %%CITATION = PHRVA,D80,123536;%%

%\cite{arXiv:1001.4444}
\bibitem{arXiv:1001.4444}
  M.~Milgrom,
  %``Matter and twin matter in bimetric MOND,''
  Mon.\ Not.\ Roy.\ Astron.\ Soc.\ {\bf 405} (2010) 1129,
  [arXiv:1001.4444 [astro-ph.CO]].
  %%CITATION = MNRAA,405,1129;%%

%\cite{arXiv:1006.3809}
\bibitem{arXiv:1006.3809}
  M.~Milgrom,
  %``Cosmological fluctuation growth in bimetric MOND,''
  Phys.\ Rev.\ D\ {\bf 82} (2010) 043523,
  [arXiv:1006.3809 [astro-ph.CO]].
  %%CITATION = PHRVA,D82,043523;%%

%\cite{arXiv:1006.5619}
\bibitem{arXiv:1006.5619}
  T.~Clifton, M.~Banados and C.~Skordis,
  %``The Parameterised Post-Newtonian Limit of Bimetric Theories of Gravity,''
  Class.\ Quant.\ Grav.\ {\bf 27} (2010) 235020,
  [arXiv:1006.5619 [gr-qc]].
  %%CITATION = CQGRD,27,235020;%%

%\cite{Eardley:1973br}
\bibitem{Eardley:1973br}
  D.~M.~Eardley, D.~L.~Lee, A.~P.~Lightman {\it et al.},
  %``Gravitational-wave observations as a tool for testing relativistic gravity,''
  Phys.\ Rev.\ Lett.\ {\bf 30} (1973) 884.

%\cite{Eardley:1974nw}
\bibitem{Eardley:1974nw}
  D.~M.~Eardley, D.~L.~Lee and A.~P.~Lightman,
  %``Gravitational-wave observations as a tool for testing relativistic gravity,''
  Phys.\ Rev.\ D {\bf 8} (1973) 3308.

%\cite{Thorne:1987af}
\bibitem{Thorne:1987af}
  K.~S.~Thorne,
  %``Gravitational Radiation,''
  in Hawking, S.W. (ed.), Israel, W. (ed.): \textit{300 years of gravitation}, 330-458,
  Cambridge University Press 1987.

%\cite{Sathyaprakash:2009xs}
\bibitem{Sathyaprakash:2009xs}
  B.~S.~Sathyaprakash and B.~F.~Schutz,
  %``Physics, Astrophysics and Cosmology with Gravitational Waves,''
  Living Rev.\ Rel.\ {\bf 12} (2009) 2,
  [arXiv:0903.0338 [gr-qc]].

%\cite{Berezhiani:2007zf}
\bibitem{Berezhiani:2007zf}
  Z.~Berezhiani, D.~Comelli, F.~Nesti and L.~Pilo,
  %``Spontaneous Lorentz Breaking and Massive Gravity,''
  Phys.\ Rev.\ Lett.\ {\bf 99} (2007) 131101,
  [hep-th/0703264].

%\cite{Berezhiani:2009kv}
\bibitem{Berezhiani:2009kv}
  Z.~Berezhiani, F.~Nesti, L.~Pilo and N.~Rossi,
  %``Gravity Modification with Yukawa-type Potential: Dark Matter and Mirror Gravity,''
  JHEP {\bf 0907} (2009) 083,
  [arXiv:0902.0144 [hep-th]].

%\cite{Will:1993ns}
\bibitem{Will:1993ns}
  C.~M.~Will,
  \textit{Theory and experiment in gravitational physics},
  Cambridge University Press 1993.

%\cite{Desai:2008vj}
\bibitem{Desai:2008vj}
  S.~Desai, E.~O.~Kahya and R.~P.~Woodard,
  %``Reduced time delay for gravitational waves with dark matter emulators,''
  Phys.\ Rev.\ D\ {\bf 77} (2008) 124041,
  [arXiv:0804.3804 [astro-ph]].

%\cite{Kahya:2010dk}
\bibitem{Kahya:2010dk}
  E.~O.~Kahya,
  %``A Useful guide for gravitational wave observers to test modified gravity models,''
  arXiv:1001.0725 [gr-qc].

%\cite{Caves:1980jn}
\bibitem{Caves:1980jn}
  C.~M.~Caves,
  %``Gravitational Radiation And The Ultimate Speed In Rosen's Bimetric Theory Of Gravity,''
  Annals Phys.\ {\bf 125} (1980) 35-52.

%\cite{Polnarev:2008tz}
\bibitem{Polnarev:2008tz}
  A.~G.~Polnarev and D.~Baskaran,
  %``The Surfing effect in the interaction of electromagnetic and gravitational waves. Limits on the speed of gravitational waves,''
  Phys.\ Rev.\ D\ {\bf 77} (2008) 124013,
  [arXiv:0802.3821 [gr-qc]].

%\cite{Baskaran:2008za}
\bibitem{Baskaran:2008za}
  D.~Baskaran, A.~G.~Polnarev, M.~S.~Pshirkov and K.~A.~Postnov,
  %``Limits on the speed of gravitational waves from pulsar timing,''
  Phys.\ Rev.\ D\ {\bf 78} (2008) 044018,
  [arXiv:0805.3103 [astro-ph]].

%\cite{Bardeen:1980kt}
\bibitem{Bardeen:1980kt}
  J.~M.~Bardeen,
  %``Gauge Invariant Cosmological Perturbations,''
  Phys.\ Rev.\ D\ {\bf 22} (1980) 1882.
  %%CITATION = PHRVA,D22,1882;%%

%\cite{Malik:2008im}
\bibitem{Malik:2008im}
  K.~A.~Malik and D.~Wands,
  %``Cosmological perturbations,''
  Phys.\ Rept.\ {\bf 475} (2009) 1,
  [arXiv:0809.4944 [astro-ph]].
  %%CITATION = PRPLC,475,1;%%

%\cite{Stewart:1990fm}
\bibitem{Stewart:1990fm}
  J.~M.~Stewart,
  %``Perturbations of Friedmann-Robertson-Walker cosmological models,''
  Class.\ Quant.\ Grav.\ {\bf 7} (1990) 1169.
  %%CITATION = CQGRD,7,1169;%%

%\cite{Newman:1961qr}
\bibitem{Newman:1961qr}
  E.~Newman and R.~Penrose,
  %``An Approach to gravitational radiation by a method of spin coefficients,''
  J.\ Math.\ Phys.\ {\bf 3} (1962) 566.

%\cite{vanderBij:1981ym}
\bibitem{vanderBij:1981ym}
  J.~J.~van der Bij, H.~van Dam and Y.~J.~Ng,
  %``The Exchange Of Massless Spin Two Particles,''
  Physica {\bf 116A} (1982) 307.
  %%CITATION = PHYSA,116A,307;%%

%\cite{Hulse:1974eb}
\bibitem{Hulse:1974eb}
  R.~A.~Hulse and J.~H.~Taylor,
  %``Discovery of a pulsar in a binary system,''
  Astrophys.\ J.\ {\bf 195} (1975) L51-L53.

%\cite{Weisberg:2010}
\bibitem{Weisberg:2010}
  J.~M.~Weisberg, D.~J.~Nice and J.~H.~Taylor,
  %``Timing Measurements of the Relativistic Binary Pulsar PSR B1913+16,''
  Astrophys.\ J.\ {\bf 722} (2010) 1030,
  [arXiv:1011.0718 [astro-ph]].

%\cite{Lorimer:2008se}
\bibitem{Lorimer:2008se}
  D.~R.~Lorimer,
  %``Binary and Millisecond Pulsars,''
  Living Rev.\ Rel.\ {\bf 11} (2008) 8,
  [arXiv:0811.0762 [astro-ph]].
\end{thebibliography}
\end{document}